\documentclass[a4paper,12pt]{amsart}
\usepackage{geometry}
\newcommand{\be}{\begin{equation}}
\newcommand{\ee}{\end{equation}}
\newcommand{\tr}{\mathrm{tr}\;}
\newcommand{\Cat}{\mbox{Cat}}
\newcommand{\F}{\mathcal{F}}

\newcommand{\Z}{\mathbb{Z}}
\newcommand{\C}{\mathbb{C}}
\newcommand{\dd}{\partial}

\newcommand{\Fun}{\mbox{Fun}}
\newcommand{\Lk}{\mbox{Lk}}
\newcommand{\dist}{\mbox{dist}}
\newtheorem{Lemma}{Lemma}
\newtheorem*{thm}{Theorem}
\newtheorem{Corollary}{Corollary}
\begin{document}
\author{P. Mn\"ev}
\title{Discrete path integral approach to the trace formula for regular graphs}
\address{PDMI RAS
\\
 27 Fontanka \\ St.-Petersburg 191023 \\ Russia }
 \email{pmnev@pdmi.ras.ru}
\begin{abstract}
We give a new proof of the trace formula for regular graphs. Our
approach is inspired by path integral approach in quantum mechanics,
and calculations are mostly combinatorial.
\end{abstract}
\maketitle

\section{Introduction}
The famous Selberg's trace formula first appeared in \cite{Selberg}.
On a compact hyperbolic surface it relates the eigenvalue spectrum
of Laplace operator  to the length spectrum of closed geodesics. A
version of this formula for finite regular graphs was obtained by
Ahumada \cite{Ahumada} (cf. also Ihara \cite{Ihara}).

Trace formulae are known to have many implications. For instance,
they can be considered as nonabelian generalizations of the Poisson
summation formula. In case of finite graphs, since one can find the
eigenvalue spectrum of Laplacian for a given graph explicitly, the
trace formula lets one find the numbers of closed geodesics of any
length (see (\ref{geodesics via eigenvalues})). In physics trace
formulae indicate the cases when semi-classical evaluation of the
path integral for state sum of a quantum free particle in some
background is exact. Selberg's formula is also known to bear much
resemblance to Riemann-Weil formula in number theory.

The original proof of the trace formula for regular graphs (\ref{sum
over geodesics}) was in the framework of ``discrete harmonic
analysis''. We propose another way to derive it, inspired by the
path integral approach in quantum mechanics \cite{Feynman}. We
consider the trace $Z_\Delta(t)=\tr e^{t\Delta}$ as a state sum of
the quantum free particle living on the graph. We rewrite it as a
sum over closed paths, which is a discrete version of the usual path
integral over loops for quantum mechanical state sum. Then we divide
the set of closed paths into classes of homotopically equivalent
paths. There is a class of contractible paths, and one homotopy
class for each closed geodesic on the graph. We explicitly calculate
the contribution of each homotopy class to the state sum, thus
rewriting it as a contribution of contractible paths plus sum over
``long'' geodesics (of nonzero length) of contributions of their
individual homotopy classes. This is analogous to the stationary
phase calculation of the path integral. Geodesics serve as
stationary points of the action in the space of loops. Homotopy
class of a geodesic serves as a neighbourhood of the stationary
point. Thus we arrive to the known trace formula for regular graph,
with a specific, physically relevant, choice of test function for
eigenvalue spectrum of Laplacian.

We wish to thank P. Zograf who inspired this work.

\section{Notations and definitions}

Let $\Gamma$ be a finite regular connected non-oriented graph with
vertices of valence $q+1\geq 2$ with no multiple edges and no edges
connecting a vertex with itself. Denote by $V(\Gamma)$ and
$E(\Gamma)$ the set of vertices of $\Gamma$ and the set of edges
respectively. Let $|\Gamma|=\#V(\Gamma)$ be the number of vertices.
Further denote the space of complex-valued functions on vertices by
$\Fun(\Gamma)=\mathbb{C}^{V(\Gamma)}$. A basis function (vector)
$|v>$ associated with vertex $v$ equals $1$ on $v$ and $0$ on the
other vertices. We further adopt the quantum-mechanical notations
and denote the transposed basis vector by $<v|=|v>^T$. We call the
set of vertices connected to $v$ by edges its link and denote it
$\Lk(v)$.

The averaging operator $T:\Fun(\Gamma)\rightarrow \Fun(\Gamma)$ acts
as follows: for $f\in \Fun(\Gamma)$ \be (Tf)(v)=\sum_{v'\in
\Lk(v)}f(v') \ee The Laplace operator $\Delta$ on $\Gamma$ is
defined by \be(\Delta f)(v)=\sum_{v'\in
\Lk(v)}f(v')-\mbox{val}(v)f(v) \ee where $\mbox{val}(v)$ is the
valence of $v$. Since we consider a regular graph $\Gamma$, $\Delta$
differs from $T$ by a multiple of identity:
$\Delta=-(q+1)\boldsymbol{1}+T$ where $\boldsymbol{1}$ is the
identity map \\ $\Fun(T)\rightarrow \Fun(T)$. The matrix of the
averaging operator is just the adjacency matrix of the graph:
~$<v'|T|v>=1$ if $v$ and $v'$ are connected by an edge and $0$
otherwise. The diagonal elements of $T$ are zero.

The physically interesting quantity is the trace of the heat kernel
(the state sum) $e^{t\Delta}$ \be
Z_\Delta(t)=\tr\exp(t\Delta)=e^{-(q+1)t}\;Z_T(t)\ee where \be
Z_T(t)=\tr\exp(tT)\ee It turns out that $Z_T(t)$ is more convenient
for our calculation than $Z_\Delta(t)$. If
$\{\lambda_j\}_{j=1}^{|\Gamma|}$ is the set of eigenvalues of $T$
then \be Z_T(t)=\sum_{j=1}^{|\Gamma|}e^{\lambda_j t}\ee If we
introduce the spectral density function for $T$
\be\rho(s)=\sum_{j=1}^{|\Gamma|}\delta(s-\lambda_j)\ee we may
express $Z_T(t)$ as a Laplace transform of $\rho(s)$: \be
Z_T(t)=\int_{-\infty}^\infty\rho(s)e^{st}ds\ee Eigenvalues
$\lambda_j$ are known to satisfy $-q-1\leq\lambda_j\leq q+1$.
Moreover, $q+1$ is always an eigenvalue, while $-q-1$ may be an
eigenvalue and may be not. If it is, then the distribution of
eigenvalues is necessarily even $\rho(-s)=\rho(s)$.

\section{Sum over paths}
Let us define a closed path of length $l$ as a sequence of vertices
$(v_1,\ldots,v_l)$ such that for every $j=1,\ldots,l$ the $v_j$ is
connected to $v_{j+1}$ by an edge (we identify $v_{l+1}$ with
$v_1$). We will usually omit the word ``closed'' in the following,
since all paths, walks, trajectories etc. will be supposed to be
closed. Denote the set of paths by $P$ and the length of a path
$p\in P$ by $|p|$. We also denote the number of closed paths of
length $l$ by $\mathbf{p}_l$. It is convenient to identify paths of
length $0$ with vertices of $\Gamma$.
\begin{Lemma}
\begin{equation}
Z_T(t)=\sum_{p\in P}\frac{t^{|p|}}{|p|!}= \sum_{l=0}^\infty
\mathbf{p}_l\;\frac{t^l}{l!}\label{sum over paths}
\end{equation}
\end{Lemma}
This expression may be viewed as a discrete version of path integral
over loops for the state sum, with $\frac{t^{|p|}}{|p|!}$ being
analogue of the measure $e^{-S}$ on loops. We give two different
explanations of (\ref{sum over paths}). The first one is more
lengthy, but done in the spirit of usual derivation of path integral
representation in quantum mechanics. The second is absolutely
straightforward and evident.
\subsection{First proof of Lemma 1}
Let us evaluate $Z_T(t)$ in the following manner:
\begin{multline}
Z_T(t)=\sum_{v\in V(\Gamma)} <v|e^{tT}|v>=
\lim_{N\rightarrow\infty}\sum_{v\in V(\Gamma)} <v|(1+\frac{t}{N}T)^N|v>= \\
=\lim_{N\rightarrow\infty}\sum_{v_1,\ldots,v_N\in V(\Gamma)}
<v_1|(1+\frac{t}{N}T)|v_N><v_N|(1+\frac{t}{N}T)|v_{N-1}>\cdots<v_2|(1+\frac{t}{N}T)|v_1>
\label{sum over walks}
\end{multline}
we are summing here over all sequences of $N$ vertices
$v_1,\ldots,v_N$. Notice that matrix elements
$<v_{i+1}|(1+\frac{t}{N}T)|v_i>$ equal $1$ if $v_{i+1}=v_i$;
$\frac{t}{N}$ if $v_{i+1}$ and $v_i$ are connected by an edge; and 0
otherwise. Let us call a walk of length $N$ a sequence of vertices
$(v_1,\ldots,v_N)$ such that each pair of successive vertices
$v_j,v_{j+1}$ are either connected by an edge or coincide. Let $W_N$
be the set of walks of length $N$. For a walk $w\in W_N$ denote the
number of values of $j$ for which $v_j$ and $v_{j+1}$ are connected
by an edge by $|w|$.

The only nonzero terms in the last line of (\ref{sum over walks})
are those with the sequence $w=(v_1,\ldots,v_N)$ being a walk. For
these terms the summand is $(t/N)^{|w|}$. Thus we have
\begin{equation}
Z_T(t)=\lim_{N\rightarrow\infty}\sum_{w\in W_N}(t/N)^{|w|}
\end{equation}
This is also a sort of discrete path integral representation for
$Z_T$. To transform it to the form (\ref{sum over paths}), we need a
projection $\pi_N:W_N\rightarrow P$ which leaves only those vertices
in a walk for which $v_j\neq v_{j+1}$, and forgets the others. For a
walk $w\in W_N$ the result of projection $\pi_N(w)$ is a path of
length $|w|$. Each path of length $l$ has $C_N^l$ walks as preimages
under $\pi_N$ ($C_N^l$ is a binomial coefficient). So \be
Z_T(t)=\lim_{N\rightarrow\infty}\sum_{p\in P}C_N^{|p|}\;(t/N)^{|p|}
\ee Using
\be\lim_{N\rightarrow\infty}\frac{C_N^{|p|}}{N^{|p|}}=\frac{1}{|p|!}\ee
we obtain (\ref{sum over paths}). $\Box$

\subsection{Second proof of Lemma 1}
One can arrive to (\ref{sum over paths}) in a more straightforward
way: we may just expand the exponent in definition of $Z_T(t)$ in a
Taylor series in variable $t$:
\be
Z_T(t)=\tr
e^{tT}=\sum_{l=0}^\infty \frac{t^l}{l!}\;\tr T^l
\ee then
\be
\tr
T^l=\sum_{v_1,\ldots,v_l\in
V(\Gamma)}<v_1|V|v_l><v_l|V|v_{l-1}>\cdots<v_2|V|v_1>
\ee
the terms
in this sum with $(v_1,\ldots,v_l)\in P$ equal 1, all the others
vanish; hence \be\tr T^l=\mathbf{p}_l\ee and we obtain (\ref{sum
over paths}). $\Box$

\section{Sum over geodesics}
We use the term ``closed trajectory'' for equivalence class of
closed paths under cyclic permutations of vertices along the path.
So a trajectory is a path with information on the starting point
forgotten. An elementary homotopy is a transformation of
trajectories of the following kind:
$$(v_1,\ldots,v_j,\ldots,v_n)\mapsto (v_1,\ldots,v_j,v',v_j,\ldots,v_n)$$
where $v'\in\Lk(v_j)$. Two trajectories are called homotopic if they
can be connected by a chain of elementary homotopies (with arrows
either forward or backward). The shortest representative in a
homotopy class is called a geodesic trajectory (or just geodesic).
An alternative definition of geodesic trajectory is as a trajectory
satisfying $v_i\neq v_{i+2}$ for all $i$. Denote the set of all
geodesics on $\Gamma$ by $G$. Two paths are called homotopic if
their trajectories are homotopic. If $\gamma=(v_1,\ldots,v_n)$ is a
trajectory of length $n$ then its $r$-th power is defined as a
trajectory of length $rn$ obtained as $\gamma$ walked around $r$
times: $\gamma^r=(v_1,\ldots,v_n,\;\ldots,\;v_1,\ldots,v_n)$. A
trajectory $\gamma$ is called primitive if it is not a (non-unit)
power of any trajectory. A geodesic trajectory with one of its
vertices chosen as a starting point is a geodesic path. A path
homotopic to path of length $0$ is called contractible. We call
geodesics of length $0$ short or trivial, and geodesics of length
$>0$ long.

We proceed now to the calculation of contribution of contractible
paths to (\ref{sum over paths}) (one may also call it the
contribution of short geodesics).

\subsection{Contribution of contractible paths}
Let us denote $\bar\Gamma$ the covering tree for $\Gamma$ and call
some point $C\in\bar\Gamma$ the center. The function $\dist$ on
vertices of the covering tree
$\dist:\bar\Gamma\rightarrow\mathbb{N}_0$ returns the minimal number
of edges one must pass to reach given vertex form the center. Any
contractible closed path on $\Gamma$ can be lifted to a closed path
on $\bar\Gamma$ (and all closed paths there are contractible, since
$\bar\Gamma$ is a tree) and we adjust the lift so that it start and
ends in $C$. Since each edge passed in one direction by a closed
path on $\bar\Gamma$ must by passed in the opposite direction, the
length of the path must be even. Denote by $P_{2k}(\bar\Gamma)$ the
set of closed paths on $\bar\Gamma$ of length $2k$ starting and
ending in $C$; we also need a subset $\tilde
P_{2k}(\bar\Gamma)\subset P_{2k}(\bar\Gamma)$ consisting of closed
paths not returning to $C$ (except the starting point and the end
point).

Recall a concept of Dyck path of length $2k$ (see e.g.
\cite{Stanley}): it is a sequence of integers $(\alpha_1, \alpha_2,
\ldots, \alpha_{2k+1})$ with $\alpha_1=\alpha_{2k+1}=0$,
$\alpha_i\geq 0$ and $\alpha_{i+1}=\alpha_i\pm 1$. We denote the set
of Dyck paths of length $2k$ as $D_{2k}$; $\# D_{2k}=\Cat_{k}$ (the
$k$-th Catalan number). There is a projection $\pi^D_{k}:\tilde
P_{2k}(\bar\Gamma)\rightarrow D_{2k-2}$. It acts as follows: \be
(v_1=C,v_2,\ldots,v_{2k},v_{2k+1}=C)\mapsto(\dist(v_2)-1,\ldots,\dist(v_{2k})-1)\ee
The number of preimages for any Dyck path under $\pi^D_{k}$ equals
$(q+1)q^{k-1}$ since there are $q+1$ choices to make the step from
$v_1=C$ to $v_2$; $q$ choices for each step, increasing $\dist$;
steps, decreasing $\dist$ are done uniquely (since for any vertex
$v\neq C$ of $\Gamma$ one edge from it leads inward, while the $q$
others lead outward). So we have obtained that \be \#\tilde
P_{2k}(\bar\Gamma)=(q+1)q^{k-1}\Cat_{k-1} \ee The generating
function for the numbers of paths $\#\tilde P_{2k}(\bar\Gamma)$ is
obtained as a simple modification of the usual generating function
for Catalan numbers: \be \F_{\tilde P}(s)=\sum_{k=1}^\infty\#\tilde
P_{2k}(\bar\Gamma)\; s^{2k}= (1+q^{-1})\frac{1-\sqrt{1-4qs^2}}{2}
\ee

For numbers of paths that may pass through the center we obtain \be
\F_{P}(s)=\sum_{k=0}^\infty\#P_{2k}(\bar\Gamma)\;
s^{2k}=\frac{1}{1-\F_{\tilde P}(s)}=
\frac{(q+1)\sqrt{1-4qs^2}-q+1}{2\;(1-(q+1)^2s^2)} \ee And hence we
obtain the contribution to $Z_T(t)$ from contractible paths:
\be
[Z_T(t)]_{con}=|\Gamma|\cdot
\sum_{k=0}^{\infty}\frac{\#P_{2k}(\bar\Gamma)}{(2k)!}t^{2k}=
|\Gamma|\cdot\frac{1}{2\pi i}\int_{A-i\infty}^{A+i\infty}ds\;
e^{st}\;\frac{1}{s}\F_P(\frac{1}{s})
\ee where real part of $A$ is
greater than real parts of all singular points of integrand, as
usual for inverse Laplace transform. The factor of $|\Gamma|$ in
front is due to the fact that a contractible path can start from any
vertex of $\Gamma$ (we remind that $|\Gamma|$ denotes the number of
vertices in $\Gamma$). Further evaluating the integral we wrap the
contour of integration around the cut $s\in [-2\sqrt q,2\sqrt q]$.
Thus we proved
\begin{Lemma} The contribution of contractible paths to (\ref{sum over
paths}) is
\begin{equation}
[Z_T(t)]_{con}=|\Gamma|\cdot\frac{q+1}{2\pi}\int_{-2\sqrt q}^{2\sqrt
q}ds\;e^{st}\;\frac{\sqrt{4q-s^2}}{(q+1)^2-s^2} \label{Z_con}
\end{equation}
\end{Lemma}
In other words, we obtained a contribution to the spectral density
of $T$ on $\Gamma$ from contractible paths:
\be
[\rho(s)]_{con}=|\Gamma|\cdot\frac{q+1}{2\pi}\frac{\sqrt{4q-s^2}}{(q+1)^2-s^2}
\ee
in the interval $s\in[-2\sqrt q,2\sqrt q]$ and 0 outside it.

\subsection{Contribution of long geodesics}
Suppose we have a (long) primitive geodesic $\gamma$ of length $l$
on $\Gamma$. To calculate the contribution of its homotopy class to
$Z_T(t)$ we need to find the number of paths of length $k$ homotopic
to $\gamma$: $\mathbf{p}_{\gamma,k}$. Since a path is a trajectory
with some point on it chosen as a start/end point,
$\mathbf{p}_{\gamma,k}=k\;\mathbf{t}_{\gamma,k}$ where
$\mathbf{t}_{\gamma,k}$ is the number of trajectories of length $k$
homotopic to $\gamma$. Note that this is not true for non-primitive
$\gamma$.

We may find the numbers $\mathbf{t}_{\gamma,k}$ using the following
combinatorial construction. If $\gamma=(v_1,\ldots,v_l)$ (with the
periodic condition $v_0=v_l$), any trajectory that can be contracted
to $\gamma$ can be represented as a closed contractible path from
$v_1$ to $v_1$ never going along the edge $v_1-v_0$; then a step
$v_1-v_2$; then a closed contractible path from $v_2$ to $v_2$,
never returning along the edge $v_2-v_1$; then step $v_2-v_3$ and so
on. All in all it is $l$ closed contractible paths with one
direction prohibited and $l$ unit steps. Thus we find the generating
function for $\mathbf{t}_{\gamma,k}$:
\be\F^{(traj.)}_\gamma(s)=\sum_{k=|\gamma|}^\infty
\mathbf{t}_{\gamma,k} s^k=s^{|\gamma|}(\hat\F_P(s))^{|\gamma|}\ee
where the superscript $(traj.)$ indicates that we are counting
trajectories, $\hat\F_P(s)$ is the generating function for the
numbers of contractible closed paths with one direction prohibited:
\be\hat\F_P(s)=\frac{1}{1-\frac{q}{q+1}\F_{\tilde
P}(s)}=\frac{1-\sqrt{1-4qs^2}}{2qs^2}\ee and hence
\be\F^{(traj.)}_\gamma(s)=\left(\frac{1-\sqrt{1-4qs^2}}{2qs}\right)^l\ee
For the numbers of paths homotopic to $\gamma$ we have
\be\F^{(paths)}_\gamma(s)=\sum_{k={|\gamma|}}^\infty
\mathbf{p}_{\gamma,k}s^k=s\frac{\dd}{\dd s}\F^{(traj.)}_\gamma(s)=
\frac{|\gamma|}{\sqrt{1-4qs^2}}\left(\frac{1-\sqrt{1-4qs^2}}{2qs}\right)^{|\gamma|}\ee

Now we would like to calculate the numbers of paths homotopic to
non-primitive geodesic $\gamma=(\gamma')^r$, that is a primitive
geodesic $\gamma'$ passed $r\geq 2$ times. It turns out that if we
carry out the scheme above in this case, every path becomes
calculated $r$ times. For any geodesic $\gamma$ denote
$\Lambda(\gamma)$ the length of the primitive geodesic $\gamma$ is
power of. If $\gamma$ is primitive itself, we set
$\Lambda(\gamma)=|\gamma|$. Thus for any geodesic we have
\be\F^{(paths)}_\gamma(s)=
\frac{\Lambda(\gamma)}{\sqrt{1-4qs^2}}\left(\frac{1-\sqrt{1-4qs^2}}{2qs}\right)^{|\gamma|}\ee

Now we have all the information to write down the contribution of a
long geodesic $\gamma$ to $Z_T(t)$:
\begin{multline}
[Z_T(t)]_{\gamma}=\frac{1}{2\pi
i}\int_{A-i\infty}^{A+i\infty}ds\;e^{st}\;
\frac{1}{s}F^{(paths)}_{\gamma}(\frac{1}{s})= \\ =\frac{1}{2\pi
i}\int_{A-i\infty}^{A+i\infty}ds\;e^{st}\;
\frac{\Lambda(\gamma)}{\sqrt{s^2-4q}}\left(\frac{s-\sqrt{s^2-4q}}{2q}\right)^{|\gamma|}
\end{multline}
The last integral reduces to the modified Bessel's function of the
first kind $I_{|\gamma|}$. So we deduced
\begin{Lemma} For every long geodesic $\gamma\in G$ the contribution of
its homotopic class in $P$ to (\ref{sum over paths}) equals
\begin{equation}
[Z_T(t)]_{\gamma}=\Lambda(\gamma)q^{-|\gamma|/2}I_{|\gamma|}(2\sqrt{q}t)
\label{Z_geod}
\end{equation}
\end{Lemma}

Collecting together (\ref{Z_con}) and (\ref{Z_geod}) we obtain the
full trace formula:
\begin{thm}[\bf Trace formula for regular graphs]
Let $\Gamma$ be a finite connected regular graph of valence $q+1\geq
2$ with $|\Gamma|$ vertices, without multiple edges and edges
connecting a vertex to itself; let $T$ be the averaging operator on
$\Gamma$ and $Z_T(t)=\tr e^{tT}$ with $t$ a complex variable; let
$G$ be the set of long closed geodesics on $\Gamma$; for each
$\gamma\in G$, $|\gamma|$ is the length of $\gamma$ and
$\Lambda(\gamma)$ is the length of the underlying primitive
geodesic: for $\gamma=(\gamma')^r$ with primitive $\gamma'$ we set
$\Lambda(\gamma)=|\gamma'|$. Then
\begin{equation}
Z_T(t)=|\Gamma|\cdot\frac{q+1}{2\pi}\int_{-2\sqrt q}^{2\sqrt
q}ds\;e^{st}\;\frac{\sqrt{4q-s^2}}{(q+1)^2-s^2}+ \sum_{\gamma\in G}
\Lambda(\gamma)q^{-|\gamma|/2}I_{|\gamma|}(2\sqrt{q}t) \label{sum
over geodesics}
\end{equation}
\end{thm}
another useful form of the same result is
\begin{equation}
Z_T(t)=|\Gamma|\cdot\frac{q+1}{2\pi}\int_{-2\sqrt q}^{2\sqrt
q}ds\;e^{st}\;\frac{\sqrt{4q-s^2}}{(q+1)^2-s^2}+ \sum_{l=3}^\infty
\mathbf{gp}_l\; q^{-l/2}I_l(2\sqrt{q}t) \label{sum over geodesics 2}
\end{equation}
where $\mathbf{gp}_l$ is the number of geodesic paths of length $l$.
To pass from (\ref{sum over geodesics}) to (\ref{sum over geodesics
2}) one must notice that the number of geodesic paths corresponding
to a given geodesic trajectory $\gamma$ is $\Lambda(\gamma)$.

We may interpret (\ref{sum over geodesics}) in terms of spectral
density $\rho(s)$
\begin{Corollary}
\begin{equation}
\rho(s)=|\Gamma|\cdot\frac{q+1}{2\pi}\;\frac{\sqrt{4q-s^2}}{(q+1)^2-s^2}\cdot\theta(4q-s^2)+
\sum_{\gamma\in
G}\Lambda(\gamma)q^{-|\gamma|/2}\frac{T_{|\gamma|}(\frac{s}{2\sqrt{q}})}{\pi\sqrt{4q-s^2}}
\cdot\theta(4q-s^2) \label{sum over geodesics for rho}
\end{equation}
where \be
T_l(x)=\cos(l\,\arccos(x))=\frac{1}{2}\left((x+\sqrt{x^2-1})^l+(x-\sqrt{x^2-1})^l\right)\ee
is the Chebyshev polynomial of the first kind of degree $l$.
\end{Corollary}
The sum on the right of  (\ref{sum over geodesics for rho}) is to be
understood in the generalized function sense (it does not exist in
the ordinary sense since numbers $\mathbf{gp}_l$ grow too fast).
Notation $\theta$ is used for unit step function. A remarkable fact
is that although each geodesic gives a smooth contribution to
$\rho(s)$ with support on the interval $s\in [-2\sqrt{q},2\sqrt{q}]$
(with singularities on the endpoints), the sum of all contributions
is a generalized function with support on eigenvalues of $T$,
scattered across a wider interval $s\in[-q-1,q+1]$.

We may invert (\ref{sum over geodesics for rho}) in a sense to
reproduce the numbers of geodesic paths $\mathbf{gp}_l$ from the
spectrum of averaging operator:
\begin{Corollary}
For $l\geq 1$
\begin{equation}
\mathbf{gp}_l=2q^{l/2}\sum_{j=1}^{|\Gamma|}T_l\left(\frac{\lambda_j}{2\sqrt{q}}\right)+
\frac{1+(-1)^l}{2}\;(q-1)\;|\Gamma| \label{geodesics via
eigenvalues}
\end{equation}
where $\lambda_j$ are eigenvalues of the averaging operator $T$.
\end{Corollary}
In particular since we know the highest eigenvalue
$\lambda_{\mbox{max}}=q+1$, we immediately get (for $q\geq 2$) from
(\ref{geodesics via eigenvalues}) the asymptotic law for numbers of
geodesic paths: $\mathbf{gp}_l\sim q^l$ as $l\rightarrow\infty$ if
$-q-1$ is not an eigenvalue. If $-q-1$ belongs to the spectrum of
$T$, the asymptotic is $\mathbf{gp}_l\sim (1+(-1)^l)\;q^l$. For
comparison the numbers of all paths behave like $\mathbf{p}_l\sim
(q+1)^l$.

\subsection{Case $q=1$}
This is a simple example where we can check (\ref{sum over
geodesics}) explicitly. The graph $\Gamma$ is necessarily a polygon
with $L\geq 3$ angles. Each long geodesic $\gamma$ is characterized
by the winding number $r\geq 1$ and its direction of movement around
the polygon: either clockwise or counterclockwise;
$\Lambda(\gamma)=L$ for all $\gamma$. The spectrum of $T$ can be
found easily: $\lambda_j=2\cos\frac{2\pi j}{L}$ and the trace
formula (\ref{sum over geodesics}) gives
\begin{equation}
Z_T(t)=\sum_{j=1}^L e^{2t\cos\frac{2\pi j}{L}}=L I_0(2t)+
2L\sum_{r=1}^\infty I_{rL}(2t)=L\sum_{r=-\infty}^\infty I_{rL}(2t)
\end{equation}
This identity can checked by Poisson resummation. In the limit
$L\rightarrow\infty$, $t=L^2\tau$ (keeping $\tau$ fixed) we recover
a special case of the modular transformation for Jacobi theta
function: \be\sum_{j=-\infty}^\infty e^{-4\pi^2j^2\tau}=
\frac{1}{\sqrt{4\pi\tau}}\sum_{r=-\infty}^\infty
e^{-\frac{r^2}{4\tau}}\ee

\subsection{Remark}
The original trace formula for regular graph in \cite{Ahumada}
(rewritten in our notations and for special case of trivial
character of the fundamental group of $\Gamma$) states that for any
function $g:\Z\rightarrow\C$ such that $g(n)=-g(n)$ for all $n\in\Z$
and \be\sum_{n=1}^\infty |g(n)|\;q^{n/2}<\infty\ee the following
holds:
\begin{equation}
\sum_{j=1}^{|\Gamma|}\hat{g}(z_j)= |\Gamma|\cdot \frac{q}{2\pi
i}\oint_{|z|=1}\hat{g}(z)\;\frac{1-z^2}{q-z^2}\;\frac{dz}{z}+
\sum_{\gamma\in G} \Lambda(\gamma) q^{-|\gamma|/2} g(|\gamma|)
\label{Ahumada}
\end{equation}
where $\hat{g}(z)=\sum_{n=-\infty}^\infty g(n)z^{-n}$ and numbers
$z_j$ are defined by eigenvalues $\lambda_j$ by equation
$\lambda=\sqrt{q}\;(z+z^{-1})$.

Formula (\ref{sum over geodesics}) follows from (\ref{Ahumada}) if
we choose $g(n)=I_n(2\sqrt{q}t)$ with corresponding
$\hat{g}(z)=e^{t\sqrt{q}(z+z^{-1})}$. Integrals representing the
contribution of contractible paths convert into one another with the
change of variables $s=\sqrt{q}\;(z+z^{-1})$.

\subsection{Remark}
The trace formula (\ref{sum over geodesics}) is actually valid for
graphs with multiple edges and loops. The condition of absence of
multiple edges and loops was chosen to simplify the combinatorial
constructions.

\end{document}